\documentstyle[multicol,aps,pre,epsf]{revtex}
\begin{document} 
\title{Uninfected random walkers in one dimension}
\author{S. J. O'Donoghue and A. J. Bray}
\address{Department of Physics and Astronomy, The University of Manchester,
M13 9PL, United Kingdom}
\date{\today}
\maketitle

\newcommand{\amod}{A \emptyset \longleftrightarrow \emptyset A}
\newcommand{\abmod}{A+B \rightarrow  \emptyset}

\begin{abstract}

We consider  a system of  unbiased diffusing walkers ($\amod$)  in one
dimension with random  initial conditions.  We investigate numerically
the relation between  the fraction of walkers $U(t)$  which have never
encountered  another  walker up  to  time  $t$,  calling such  walkers
``uninfected'' and the fraction of  sites $P(t)$ which have never been
visited by a  diffusing particle.  We extend our  study to include the
$\abmod$  diffusion-limited  reaction  in  one-dimension,  with  equal
initial densities  of $A$ and $B$  particles distributed homogeneously
at $t=0$.  We find $U(t)  \simeq [P(t)]^\gamma$, with $\gamma \simeq 
1.39$, in both models, though there is evidence that a smaller value of 
$\gamma$ is required for $t \to \infty$.    

\end{abstract}

\begin{multicols}{2}

\section{Introduction}

Random  walks  model a  host  of  phenomena  and find  application  in
virtually  all areas of  physics \cite{1,2,3,4,5}.   In this  paper we
consider  a one-dimensional (1D)  system of  non-interacting, unbiased
diffusing walkers $\amod$,  with random initial conditions.  Recently,
we  have studied  the  \textit{persistence} properties  of this  model
\cite{6}. The  persistence probability $P(t)$ has  been widely studied
in  recent years \cite{6,7,8,9,10,11,12,13,14,15,16,17,18,19}.  In the
present context $P(t)$ is defined  as the fraction of sites which have
never been visited by a diffusing particle up to time $t$.  Such sites
are termed ``persistent sites''. We  showed \cite{6} that, in the case
of the $\amod$ processes,
\begin{equation}
P(t) = \exp[-(16\rho_{0}^{2}Dt/\pi)^{1/2}]\ ,
\label{eq1}
\end{equation}
where  $\rho_{0}$ is  the particle  density and  $D$ is  the diffusion
constant. The fraction of unvisited sites in this model can equally be
thought  of as  the surviving  fraction of  immobile particles  in the
presence of randomly diffusing traps.  At $t=0$, consider each site to
contain a static particle, $A$, so that $\rho_{A}(0) = 1$. Then for $t
> 0$, let particles of another  type, $B$, randomly distributed on the
lattice with density $\rho_0$, diffuse though the system  with the $B$
particles acting as  perfect traps for the $A$  particles so that $A+B
\rightarrow  B$. This  model  is just  the well-studied  ``scavenger''
model of  Redner and Kang \cite{20,21,22}.  The  fraction of unvisited
sites is then just $\rho_{A}(t)$.

The focus of this paper is  to explore the relation between $P(t)$ and
the  fraction of walkers  $U(t)$ that  have never  encountered another
walker up to time $t$.  To facilitate our discussion, we introduce the
following    definition.    We    consider   all    walkers    to   be
\textit{uninfected}   at  time  $t=0$.   The  walkers   are,  however,
considered to be mutually infectious, so  that for $t > 0$ any contact
between walkers leads to mutual contamination.  We therefore define an
uninfected walker to be one  which has never encountered (occupied the
same  site as)  another walker.  We can  then refer  to $U(t)$  as the
fraction of uninfected walkers. The fraction of uninfected walkers is,
in fact,  isomorphic to the  survival probability of $A$  particles in
the reaction $A+B \rightarrow B$, with equal mobility for both species
\cite{20,23,24,25}. At late times, one expects the asymptotics of this
model  to be  identical  to those  of  the $\abmod$  diffusion-limited
process with $\rho_{A}(0) \ll \rho_{B}(0)$. For this process it was 
proved rigorously by Bramson and Lebowitz \cite{24} that 
\begin{equation}
\rho_{A}(t) \sim \exp(- \lambda t^{1/2})
\label{eq2}
\end{equation}
in  one-dimension,  with  unknown  constant $\lambda$.   Therefore  we
expect identical  asymptotic behavior for $U(t)$. The  kinetics of the
two  species annihilation  model has  received  considerable attention
over   the    past   two    decades   and   is    largely   understood
\cite{5,23,24,26,27,28,29,30}.  It is  well appreciated  that  in this
model, there  exists an upper critical dimension  $d_{c}$, below which
spatial fluctuations in the initial distribution of the reactants play
a  significant   role  in  the   evolution  of  the  density   of  the
particles. This dependence on the microscopic fluctuations invalidates
traditional  approaches  such as  the  mean-field approximation  which
yields $\rho(t)  \sim t^{-1}$.   It was first  shown by  Toussaint and
Wilczek \cite{26}  that, for  equal initial densities  ($\rho_{A}(0) =
\rho_{B}(0)$) of randomly diffusing  $A$ and $B$ particles, the walker
density   decays   anomalously,   according   to   $\rho_{A}(t)   \sim
\sqrt{\rho_{A}(0)}t^{-d/4}$ for  $d \le d_{c} = 4$,  assuming that the
particles are distributed homogeneously at  $t=0$. In the case of this
model we  argued heuristically, on the  basis of a  toy model \cite{6}
that,
\begin{equation}
P(t) \sim \exp[-C(\rho_{0}^{2} D t) ^{1/4}]
\label{eq3}
\end{equation}
where $C$ is a constant and $\rho_{0} = \rho_{A}(0) +\rho_{B}(0)$.  We
extend  our study  to  include  an investigation  of  the fraction  of
uninfected walkers in this model in one-dimension.

The fraction of uninfected walkers, $U(t)$, is related to the fraction
of persistent  sites, $P(t)$, in the following  manner.  In evaluating
$P(t)$,  one  considers  particles  with diffusion  constant  $D$  and
addresses the probability that a given site $x$ has never been visited
by any  walker up to time  $t$.  Equivalently, one  could consider the
site  $x$ to  contain a  particle with  diffusion constant  $D'=0$ and
address the  probability that  the particle remains  uninfected (i.e.\
unvisited).   However, if the  particle initially  at position  $x$ is
given a diffusion constant equal to that of the other particles in the
system i.e.\ $D'=D$,  then the probability that it  ``persists'' up to
time $t$ is the probability that it remains uninfected.  This suggests
that the number of uninfected particles is a function of the diffusion
constants in the problem,  with the limiting case $D'=0$ corresponding
to  the standard  persistence  problem.  Clearly,  in  order to  avoid
contamination, remaining stationary is  a more effective strategy than
diffusing. Therefore,  we anticipate  that the fraction  of uninfected
walkers  decays   more  rapidly   than  the  fraction   of  uninfected
(unvisited, i.e.\  persistent) sites. This  faster decay of  $U(t)$ is
clearly observed in the simulations.

Given  the  similar  decay  of  $P(t)$ in  the  $\amod$  and  $\abmod$
processes, these models make a sensible combination to study together.
We discover that  in both cases, $U(t) \sim  [P(t)]^\gamma$ where $1 <
\gamma  \lesssim 1.39$.  In   fact  the  relation  $U(t)  =
k[P(t)]^\gamma$,  with  $k$   close  to  unity  and  $\gamma=1.39(1)$,
describes the  data rather  accurately in both  models over  the whole
range of times studied, although  there are indications in both models 
that  the  asymptotic  value of  $\gamma$ is  smaller than $1.39$.  We 
discuss the $\amod$ model  in Sec.\  II and follow  with the  $\abmod$  
reaction in Sec.\ III.  We conclude in Sec.\  IV with a discussion and  
summary of the results.

\section{The $\amod$ model}

We consider a system  of non-interacting, unbiased, diffusing walkers,
with random  initial conditions.  Our numerics are  performed on  a 1D
lattice of  size $L=10^{7}$ with  lattice constant $a=1$  and periodic
boundaries. At  $t=0$, $N$ random walkers are  randomly distributed on
the lattice with  a maximum of one particle per  site so that $U(0)=1$
and $P(0) = 1 - \rho_{0}$,  where $\rho_{0} = \rho(0) = N/L$. Clearly,
$P(0) \neq 1$  is a consequence of our lattice  description and in the
continuum  limit $\rho_{0}  \rightarrow  0$, where  the mean  distance
between  walkers  is  much  larger  than the  lattice  spacing,  $P(0)
\rightarrow  1$.  Therefore,  in   order  to  place  the  fraction  of
persistent sites  and the fraction  of uninfected walkers on  an equal
footing in our numerical study, we define,
\begin{equation} 
p(t) = \frac{P(t)}{P(0)}
\label{eq4}
\end{equation}
so that both $U(0)$ and $p(0)$  are normalized to 1. We now use $p(t)$
in the description of our numerical results.

Our model is  updated using the direct method  \cite{31} i.e.\ at each
computational step,  a particle is  picked at random and  shifted with
probability $D=1/2$ to a neighboring  site, where $D$ is the diffusion
constant.  For  $t>0$, any site which  is visited by  a walker becomes
nonpersistent for all remaining time $t$. Mutual infection occurs when
two  or more particles  simultaneously occupy  a single  lattice site.
For each jump  made by a particle, time $t$  is increased by $dt=1/N$,
where $N$ is  the current number of particles in  the system (which is
constant  in the  present model).  Our results  are averaged  over 100
independent runs.

Dimensional  analysis  demands  that,  on  a  \textit{continuum},  the
fraction  of  unvisited  sites  is  a function  of  the  dimensionless
combination     of     parameters     in    the     problem,     i.e.\
$P(t)=f(\rho_{0}^{2}Dt)$.  Similarly,  for the fraction  of uninfected
walkers,   $U(t)=g(\rho_{0}^{2}Dt)$,   where   $g$   is   some   other
function. However, given that  our numerical simulations take place on
a discrete  lattice, we expect the  scaling of $P(t)$ and  $U(t)$ as a
function  of $\rho_{0}^{2}Dt$  to  become strictly  true  only in  the
continuum  limit, $\rho_{0}  \rightarrow 0$.  In order  to approximate
this  limit  as  closely  as  possible, we  choose  as  our  densities
$\rho_{0} = 0.007$, 0.005 and 0.003, so that $\rho_{0} \ll 1$.

The  dimensionless combination  of parameters,  $\rho_{0}^{2}Dt$, also
defines  a natural characteristic  time scale  in the  problem, namely
$t^* =  1/\rho_0^2D$. Therefore one  expects to observe  the asymptotic
behavior  of  the system  only  after  a time  $t  \gg  t^*$ has  been
reached. However,  as far as the persistent  properties are concerned,
Eq.\ (\ref{eq1}) is  an exact relation in the  continuum limit for all
$t$.  Therefore,  we expect  to observe $P(t)$  approaching asymptopia
very  quickly  in  our  numerical  data.   Clearly,  in  the  discrete
description  of the model  used in  our simulations,  Eq.\ (\ref{eq1})
breaks down at  the earliest times since, for  very small $t$, lattice
effects play  a dominant role. Eq.\ (\ref{eq2}),  however, which holds
for  $U(t)$  is  a  truly  asymptotic  relation.   Unfortunately,  the
extremely fast decay  of $U(t)$ leads to small-number  effects at very
early  times in  our simulations  and we  find the  limit $t  \gg t^*$
unattainable with  good statistics.  Therefore, we  limit ourselves to
studying the regime up to $t_{max} \simeq t^*/2$.

\begin{figure}
\narrowtext \centerline{\epsfxsize\columnwidth\epsfbox{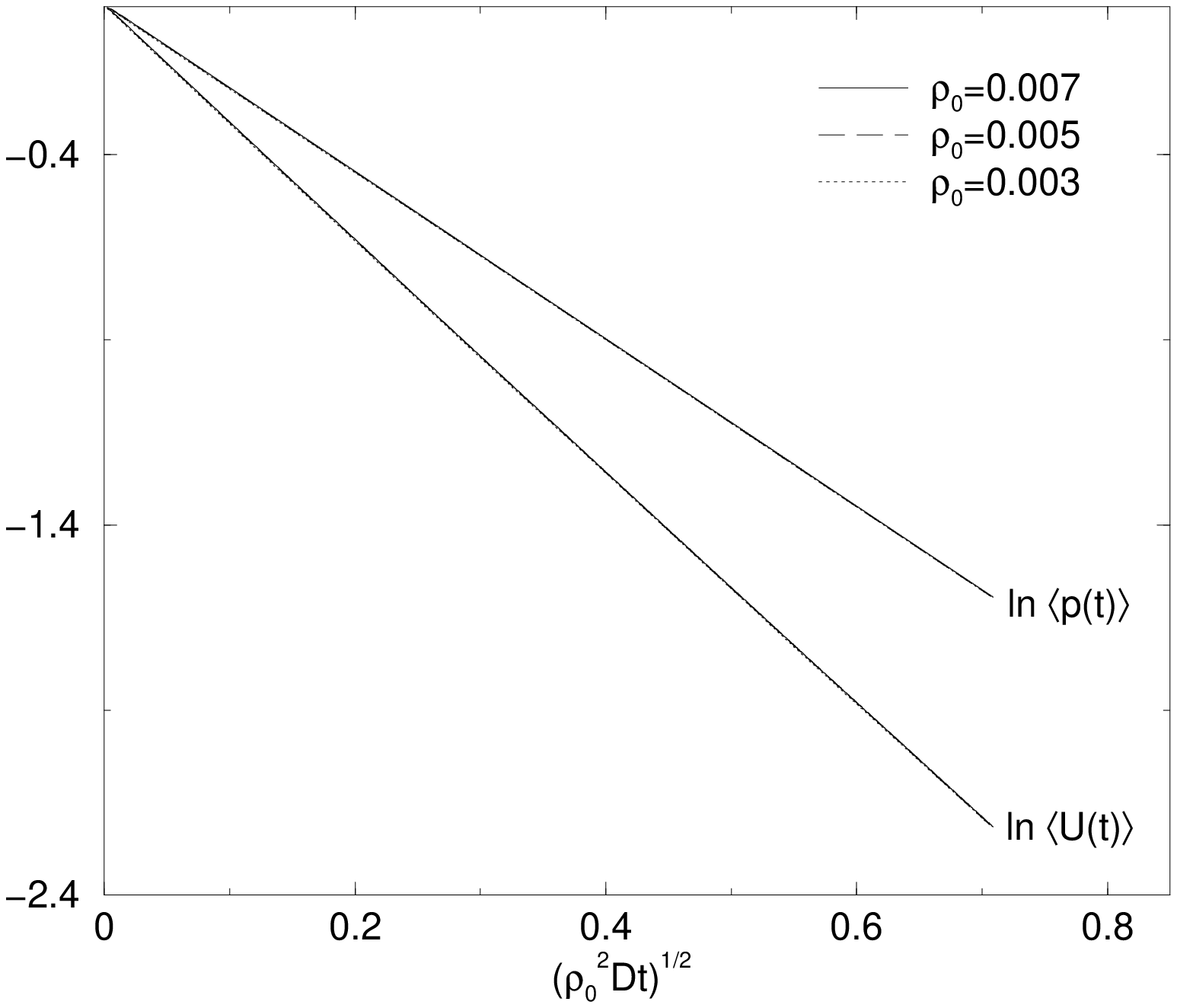}}
\caption{Log-linear  plot  of  the  fraction  of  unvisited  sites  
$p(t)$, and  the  fraction   of  uninfected  walkers  $U(t)$,  as   
a  function  of $(\rho_{0}^{2}Dt)^{1/2}$, for the $\amod$ model.}
\label{fig1}
\end{figure}

The  fraction  of unvisited  sites  decays,  in  the continuum  limit,
exactly as the stretched exponential form, Eq.\ (1) \cite{6}. In order
to make a direct comparison between the fraction of persistent  sites,
$p(t)$,  and the fraction  of uninfected  walkers, $U(t)$,  we present
these two quantities on a  log-linear plot in Fig.\ \ref{fig1} , using
$(\rho_0^2Dt)^{1/2}$ as abscissa.

\begin{figure}
\narrowtext \centerline{\epsfxsize\columnwidth\epsfbox{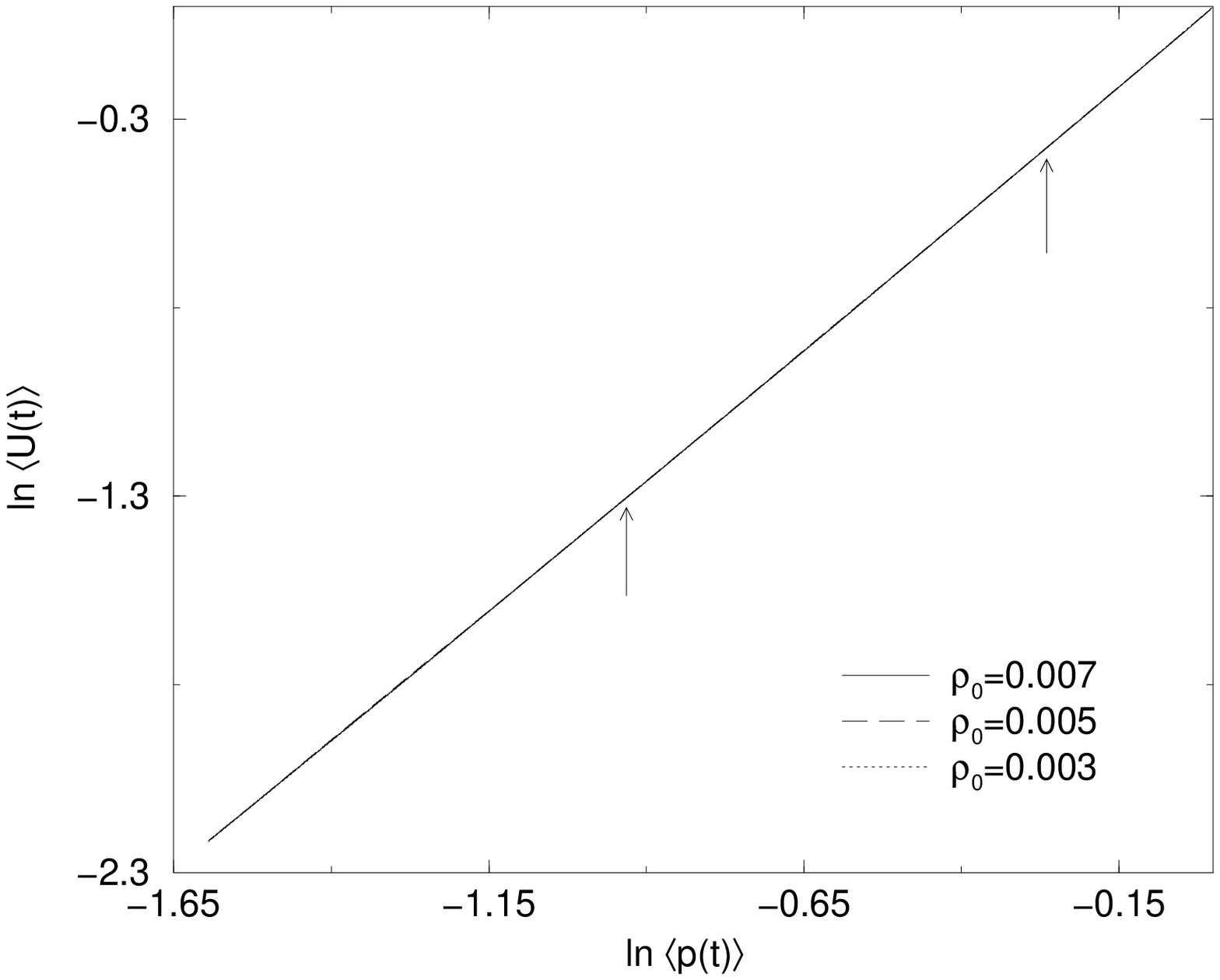}}
\caption{Log-log plot  of the fraction of  uninfected walkers, $U(t)$,
against  the  fraction of  unvisited  sites,  $p(t)$,  in the  $\amod$
model.}
\label{fig2}
\end{figure}

\noindent  Fig.\ \ref{fig1}  shows clearly  that the  data scale  as a
function  of  the   dimensionless  quantity  $\rho_{0}^{2}Dt$.  $U(t)$
exhibits behavior  characteristic of  Eq.\ (\ref{eq2}), with  a faster
decay rate than for the  fraction of persistent sites i.e.\ $\lambda >
(16 \rho_{0}^{2}D/ \pi)^{1/2}$. The data seem consistent with
\begin{equation}
U(t) \simeq K[p(t)]^\gamma\ , 
\label{eq5}
\end{equation}
with $K \simeq 1$ and $\gamma  >1$.  The log-log plot of $U(t) \textit
{  vs.  }  p(t)$  in Fig.\  \ref{fig2} indeed  shows an  almost linear
relationship between $\ln U(t)$ and  $\ln p(t)$.  In order to evaluate
$\gamma$ and $K$ we performed a linear regression on the data in Fig.\
\ref{fig2}  in the region  indicated by  the arrows,  thereby avoiding
initial transients  associated with  our lattice description,  and the
onset of statistical fluctuations  between different runs of the data.
We summarize our results for $\gamma$ and $K$ in Table\ 1.

\bigskip
\begin{center}
\begin{tabular}{|c|c|c|}
\hline \hline
\hspace{0.5cm} $\rho_{0}$ \hspace{0.5cm} & \hspace{0.5cm} $\gamma$
\hspace{0.5cm} & \hspace{0.5cm} $K$ \hspace{0.5cm} \\ \hline
\hspace{0.5cm} 0.007 \hspace{0.5cm} & \hspace{0.5cm} 1.391(5)
\hspace{0.5cm} & \hspace{0.5cm} 0.990(3) \hspace{0.5cm} \\
\hspace{0.5cm} 0.005 \hspace{0.5cm} & \hspace{0.5cm} 1.391(5)
\hspace{0.5cm} & \hspace{0.5cm} 0.991(3) \hspace{0.5cm} \\
\hspace{0.5cm} 0.003 \hspace{0.5cm} & \hspace{0.5cm} 1.391(5)
\hspace{0.5cm}  &  \hspace{0.5cm}  0.991(3) \hspace{0.5cm}  \\  \hline
\hline
\end{tabular}
\end{center}

\begin{small}
TABLE  1.  Numerical  values of  $\gamma$  and $K$,  where $U(t)\simeq
K[p(t)]^\gamma$, for various $\rho_{0}$, in the $\amod$ model.
\end{small}

\bigskip

\begin{figure}
\narrowtext \centerline{\epsfxsize\columnwidth\epsfbox{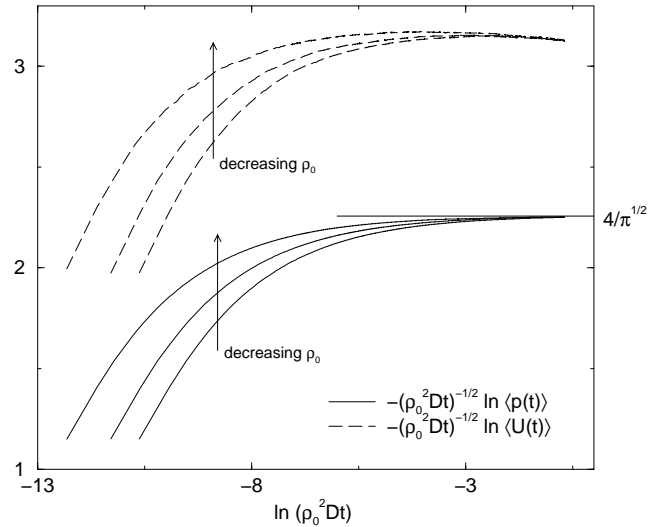}}
\caption{$-(\rho_{0}^{2}Dt)^{-1/2}         \ln        p(t)$        and
$-(\rho_{0}^{2}Dt)^{-1/2}    \ln    U(t)$    plotted   against    $\ln
(\rho_{0}^{2}Dt)$ for $\rho_{0}=0.007,0.005,0.003$.}
\label{fig3}
\end{figure}

The data are  sufficiently good to merit closer  inspection.  To check
how   far  our  numerics   probe  the   asymptotic  regime,   we  plot
$-(\rho_{0}^{2}Dt)^{-1/2} \ln  p(t)$ and $-(\rho_{0}^{2}Dt)^{-1/2} \ln
U(t)$ against $\ln (\rho_{0}^{2}Dt)$  in Fig.\ \ref{fig3}.  The nature
of  this plot  greatly expands  the  early time  regime.  Whereas  the
data-collapse   for  different   densities  is   excellent   in  Fig.\
\ref{fig1},  a systematic  splitting of  the curves  as a  function of
$\rho_{0}$ is now  clearly evident.  This feature is  a consequence of
the discreteness  of the  lattice. In fact  the behavior of  $p(t)$ is
well  described,  for  all   $t$,  by  an  exact  lattice  calculation
\cite{Blythe} which predicts
\begin{equation}
p(t) = \exp[-\rho_0a\,f(Dt/a^2)]\ ,
\label{eq5a}
\end{equation} 
where $a$ is the lattice spacing. The function $f(x)$ has the limiting
behavior $f(x) \to  2x$ for $x \to 0$ and  $f(x) \to (16 x/\pi)^{1/2}$
for $x \to \infty$, the latter reproducing the continuum limit result,
Eq.\  (\ref{eq1}). Indeed,  the structure  of Eq.\  (\ref{eq5a}) shows
that the continuum  limit, $a \to 0$ at fixed  $t$, and the asymptotic
limit, $t  \to \infty$  at fixed $a$  are the same.  Eq.\ (\ref{eq5a})
also shows that the quantity  $-(\rho_0^2 Dt)^{-1/2} \ln p(t)$ is, for
fixed  $a$ and  $D$,  a function  of  $t$ only,  i.e.\ independent  of
$\rho_0$.  For small  $t$ it behaves as $t^{1/2}$,  accounting for the
small-$t$   behavior   in   Fig.\  (\ref{fig3}).    Plotting   against
$\ln(\rho_0^2 Dt)$ displaces the curves for different densities in the
horizontal direction,  as is  clear in the  figure.  The  whole region
where the curves are split  in Fig.\ \ref{fig3} corresponds to the top
left-hand corner of Fig.\ \ref{fig1}.  If the data are plotted against
$\ln t$ in Fig.\ (\ref{fig3}), instead of $\ln(\rho_0^2 Dt)$, the data
for  different densities recollapse,  but the  onset of  the continuum
limit at late times not as clear.

For large $t$,  where the curves for the  different densities converge
in these  scaled variables, the persistence data  clearly approach the
limiting value  $4 / \sqrt{\pi}$, consistent with  the continuum limit
displayed  in Eq.\  (\ref{eq1}). By  contrast, Fig.\  \ref{fig3} shows
that all  of our  data for $U(t)$  is merely preasymptotic.   The data
reach  a maximum  and then  appear  to monotonically  decrease in  the
region where the different  curves converge (the continuum limit). Our
results  are  consistent with  the  recent  simulations  of Mehra  and
Grassberger  \cite{25} who probe  $U(t)$ to  much smaller  values than
achieved  here. The  slow approach  to  asymptopia for  $U(t)$ is  not
understood  and, although  our numerics  seem roughly  consistent with
Eq.\ (\ref{eq2}), a numerical determination of $\lambda$ appears to be
a hard problem. In the light of Fig.\ \ref{fig3}, Fig.\ \ref{fig2} and
its associated  results (Table\  1) are somewhat  misleading. Clearly,
over  the times  covered by  our simulations,  $\gamma$ can  not  be a
constant since $\lambda$ is not.

\begin{figure}
\narrowtext \centerline{\epsfxsize\columnwidth\epsfbox{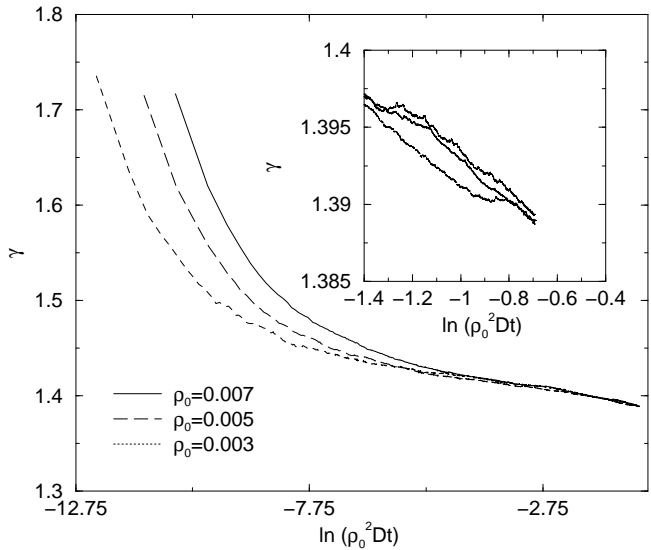}}
\caption{Plot of  $\gamma$ \textit  {vs.}  $\ln (\rho_0^2  Dt)$, where
$\gamma = \ln U(t)/\ln p(t)$, for the $\amod$ model. Inset: blow-up of
the right-hand part of the plot.}
\label{fig4}
\end{figure}

In Fig.\  \ref{fig4} we plot  the effective exponent $\gamma(t)  = \ln
U(t)/\ln p(t)$  directly as  a function of  time.  The data  show that
$\gamma$ decreases   monotonically with  time even in  the ``continuum
regime'' where  the curves for  different densities converge.   For $t
\rightarrow  \infty$,  $\gamma$  must  tend  to  a  constant  if  Eq.\
(\ref{eq2}) is correct. We note  that since the fraction of uninfected
walkers decreases more rapidly  than the fraction of persistent sites,
$\gamma > 1$.   Therefore, if the trend in the  data continues, we can
bound $\gamma$  by $1  < \gamma <  1.39$: Asymptotically,  $U(t) \sim
[P(t)]^{\gamma}$ for some fixed $\gamma$ where $1 < \gamma < 1.39$.

Let us  consider once again the  probability of being  uninfected of a
particle  at site  $x$, with  diffusion constant  $D'$,  and diffusing
particles to its left and  right with diffusion constant $D$. Then, we
suggest that the fraction of uninfected walkers can be expressed as,
\begin{equation}
U(t) \sim \exp[-(\rho_{0}^{2}Dt)^{1/2}F(D'/D)]
\label{eq6}
\end{equation} 
for  large $t$,  consistent with  the result  of Bramson  and Lebowitz
(Eq.\ (\ref{eq2})).  $F(x)$  is some unknown function of  the ratio of
the  diffusion constants  in the  system. For  $D' =  0$,  the problem
reduces to  the familiar study  of unvisited sites. Therefore  $F(0) =
(16/\pi)^{1/2}$  and $U(t) =  P(t)$.  The  case $D'=D$  corresponds to
equally  mobile  particles,  and  the  particle at  site  $x$  becomes
infected more quickly than if it had remained stationary. In this case
$F(1) = \gamma(16/\pi)^{1/2}$  giving $U(t) \sim [P(t)]^\gamma$ where,
according  to our numerics,  $1 <  \gamma <  1.39$.  Although  we have
explored the  values $D'=0$ and  $D'=D$ only, we conjecture  that Eq.\
(\ref{eq6}) holds for any ratio of the diffusion constants.

\section{The $\abmod$ Model}

We now address the fraction  of uninfected walkers, $U(t)$, in the two
species  annihilation model.  Our  simulations are  performed on  a 1D
lattice  of size  $L=10^{7}$  with periodic  boundary conditions.   At
$t=0$,  exactly equal  numbers,  $N_{A}(0)=N_{B}(0)$, of  $A$ and  $B$
particles are  randomly distributed on  the lattice with a  maximum of
one  particle per  site, such  that  at large  scales, both  densities
$\rho_{A}(0)$ and $\rho_{B}(0)$  are initially homogeneous.  We define
$\rho_{A}(0)=\rho_{B}(0)=N_{A}(0)/L$                                and
$\rho_{0}=\rho_{A}(0)+\rho_{B}(0)$.  Both  species are also  given the
same diffusion constant $D_{A}=D_{B}=D=1/2$. Our model is then updated
in the  same manner as  for the $\amod$  model described in  Sec.\ II,
i.e.\  $dt=1/N(t)$, where  $N(t)  = N_{A}(t)+N_{B}(t)$,  is the  total
current number of particles in  the system.  Infection occurs when two
or  more particles  occupy a  single lattice  site, but  we  impose an
instantaneous reaction  ($\abmod$) so that each  lattice contains only
one type of particle.

It is  well known that  in the two  species annihilation model  in one
dimension,  there is  a effective  repulsion between  the $A$  and $B$
particles  that   favors  segregation  into   single  species  domains
\cite{5,26,30}. This domain coarsening  leads to early time transients
for both  $P(t)$ and  $U(t)$.  Although using  a high  initial density
serves  to  accelerate the  systems  progression  into the  asymptotic
regime, in doing so, not only  is the continuum limit obscured, but we
also  find that the  extremely fast  decay of  both $P(t)$  and $U(t)$
leads to small-number effects at very early times. Therefore we choose
$\rho_{0} \ll 1$. For consistency,  we select our parameters to be the
same  as those  studied  in  the $\amod$  process,  i.e.\ $\rho_{0}  =
0.007$, 0.005  and 0.003, and $\rho_{0}^{2}Dt_{max}  \simeq 0.5$ where
$t_{max}$ is the maximum number  of time steps in the simulations. All
the results are averaged over 100 runs.

\begin{figure}
\narrowtext \centerline{\epsfxsize\columnwidth\epsfbox{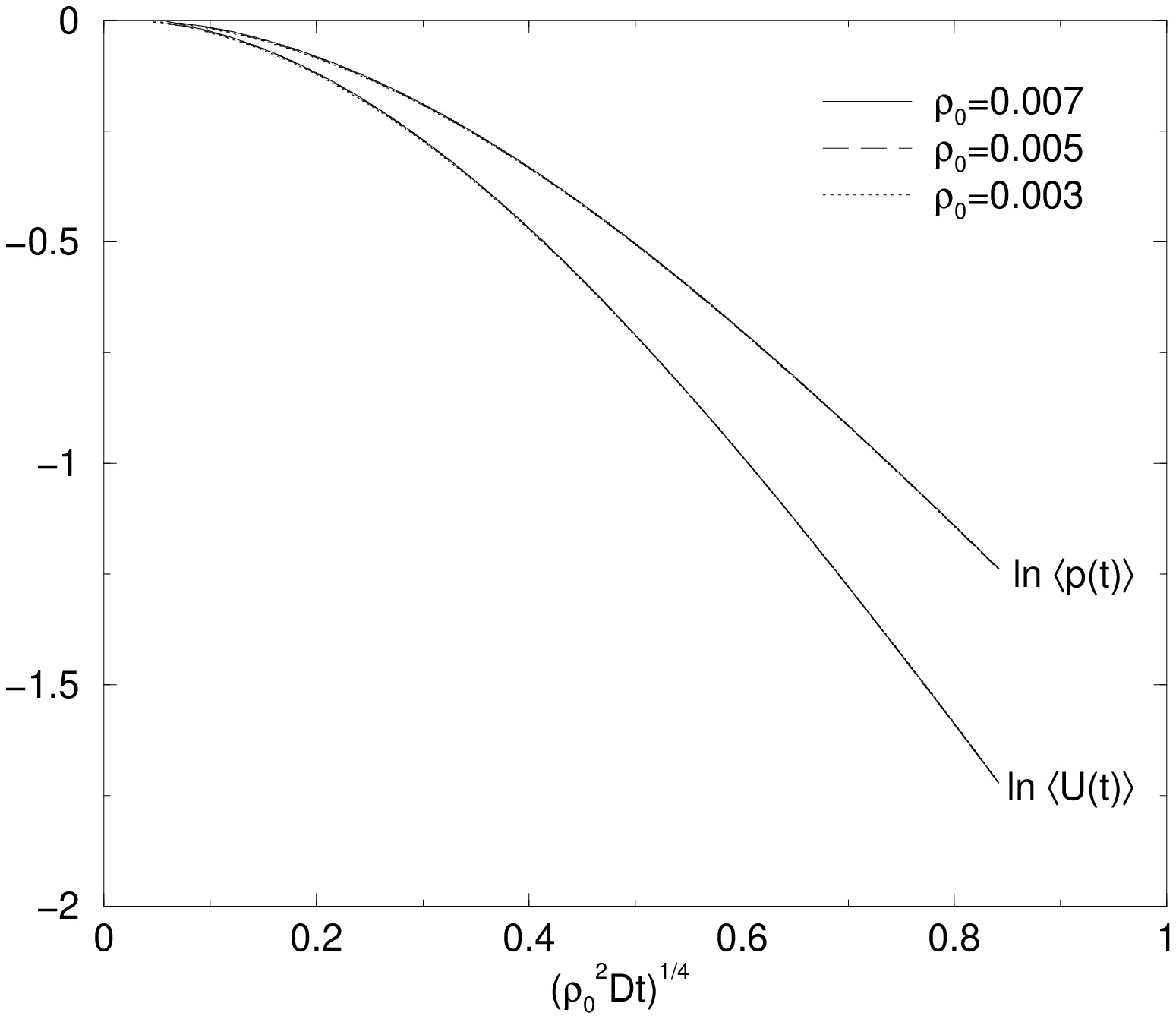}}
\caption{Log-linear plot  of the  fraction of unvisited  sites $p(t)$,
and  the fraction  of  uninfected  walkers $U(t)$,  as  a function  of
$\rho_{0}^{2}Dt$, for the $\abmod$ model.}
\label{fig5}
\end{figure}

\begin{figure}
\narrowtext \centerline{\epsfxsize\columnwidth\epsfbox{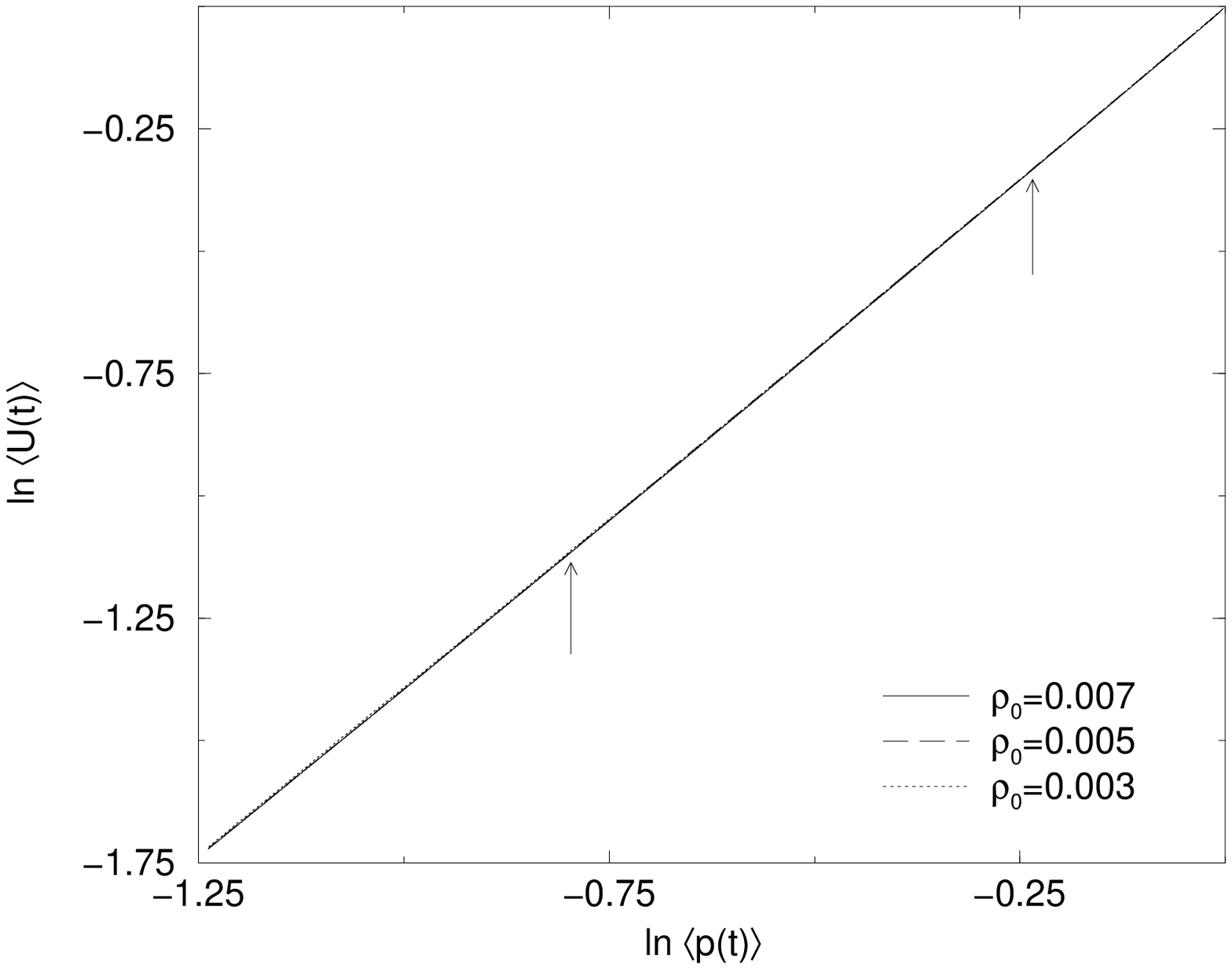}}
\caption{Log-log plot  of the fraction of  uninfected walkers, $U(t)$,
against  the  fraction of  unvisited  sites, $p(t)$,  in the  $\abmod$
model.}
\label{fig6}
\end{figure}

The  format of  our analysis  is  much the  same as  in the  preceding
section.   We   observe  that   an  algebraic  relation,   $U(t)  \sim
[p(t)]^\delta$, between $U(t)$ and  $p(t)$ also holds approximately in
this  case over the  range of  times studied.   Surprisingly, $\delta$
seems to have a similar value  to $\gamma$ in the $\amod$ model, i.e.\
$1 < \delta \lesssim 1.39$. For the two-species annihilation
process  we suggested  in our  earlier paper,  on the  basis of  a toy
model,  that the  fraction  of unvisited  sites decays  asymptotically
according  to   the  stretched  exponential   form,  Eq.\  (\ref{eq3})
\cite{6}.   In  the manner  of  Fig.\  \ref{fig1},  to make  a  direct
comparison between  the fraction of  persistent sites $P(t)$,  and the
fraction of uninfected walkers $U(t)$, we present these two quantities
on a log-linear  plot in Fig.\ \ref{fig5}, this  time using $(\rho_0^2
Dt)^{1/4}$  as  abscissa, as  suggested  by  Eq.\ (\ref{eq3}).   Fig.\
\ref{fig5} shows clearly that the data scale excellently as a function
of  the  dimensionless   quantity  $\rho_{0}^{2}Dt$.   We  appreciate,
however, that  the stretched exponential behavior  of Eq.\ (\ref{eq3})
is not particularly  well realized. This may be  due to the prevalence
of large initial transients  (we presented more convincing evidence of
the  exponential   behavior  in  our  earlier   paper  \cite{6}).   In
particular, the asymptotic behavior represented in Eq.\ (\ref{eq3}) is
only  expected   for  large   values  of  $\rho_0^2Dt$,   unlike  Eq.\
(\ref{eq1}) for $P(t)$,  which we have shown holds for  all $t$ on the
continuum \cite{6}. Nonetheless, even in the regime we have studied it
is clear from Fig.\ \ref{fig5} that the fraction of uninfected walkers
has a very  similar form of decay to the  fraction of unvisited sites,
but with a faster decay  rate. In Fig.\ \ref{fig6}, therefore, we plot
$\ln  U(t)$  against $\ln  p(t)$  to see  whether  there  is a  simple
relationship between $U(t)$ and $p(t)$.

The  approximate linearity of  the data  in Fig.\  \ref{fig6} suggests
$U(t)  \sim  k[p(t)]^{\delta}$.  To  determine  $\delta$  and $k$,  we
performed a linear  regression on the data in  the region indicated by
the arrows,  thereby avoiding  initial transients associated  with our
lattice description, and the onset of statistical fluctuations between
different runs of the data.  We summarize our results for $\delta$ and
$k$ in  Table\ 2. Notice that  $\delta$ and $k$  agree remarkably well
with $\gamma$ and $K$ recorded in Table\ 1 for the $\amod$ process.

\bigskip
\begin{center}
\begin{tabular}{|c|c|c|}
\hline \hline
\hspace{0.5cm} $\rho_{0}$ \hspace{0.5cm} & \hspace{0.5cm} $\delta$
\hspace{0.5cm} & \hspace{0.5cm} $k$ \hspace{0.5cm} \\ \hline
\hspace{0.5cm} 0.007 \hspace{0.5cm} & \hspace{0.5cm} 1.390(5)
\hspace{0.5cm} & \hspace{0.5cm} 0.990(3) \hspace{0.5cm} \\
\hspace{0.5cm} 0.005 \hspace{0.5cm} & \hspace{0.5cm} 1.390(5)
\hspace{0.5cm} & \hspace{0.5cm} 0.993(3) \hspace{0.5cm} \\
\hspace{0.5cm} 0.003 \hspace{0.5cm} & \hspace{0.5cm} 1.386(5)
\hspace{0.5cm}  &  \hspace{0.5cm}  0.991(3) \hspace{0.5cm}  \\  \hline
\hline
\end{tabular}
\end{center}

\begin{small}
TABLE  2.  Numerical  values of  $\delta$  and $k$,  where $U(t) \simeq
k[p(t)]^\delta$, for various $\rho_{0}$, in the $\abmod$ model.
\end{small}

\bigskip

\begin{figure}
\narrowtext \centerline{\epsfxsize\columnwidth\epsfbox{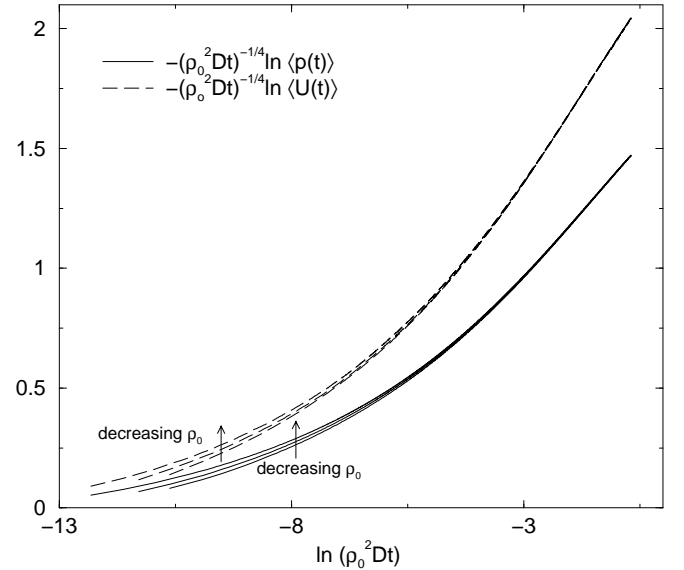}}
\caption{$-(\rho_{0}^{2}Dt)^{-1/4}         \ln        p(t)$        and
$-(\rho_{0}^{2}Dt)^{-1/4}    \ln    U(t)$    plotted   against    $\ln
(\rho_{0}^{2}Dt)$ for $\rho_{0}=0.007,0.005,0.003$.}
\label{fig7}
\end{figure}

In analogy  to Fig.\ \ref{fig3}, we  plot $(\rho_{0}^{2}Dt)^{-1/4} \ln
p(t)$   and   $(\rho_{0}^{2}Dt)^{-1/4}    \ln   U(t)$   against   $\ln
(\rho_{0}^{2}Dt)$ in Fig.  \ref{fig7}.

\begin{figure}
\narrowtext \centerline{\epsfxsize\columnwidth\epsfbox{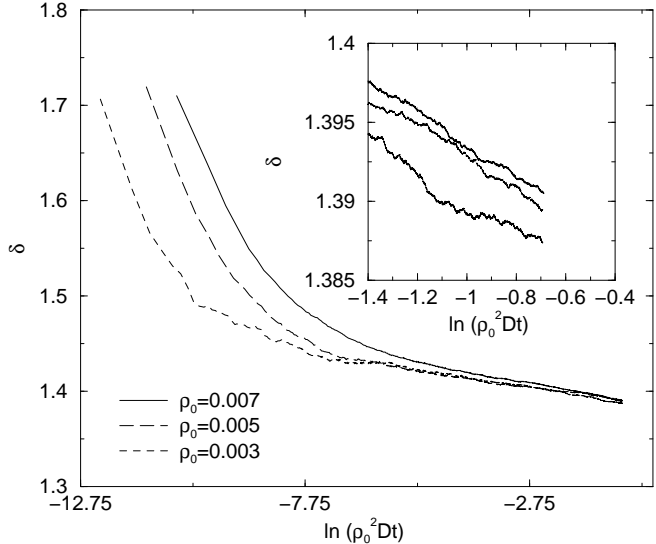}}
\caption{Plot of $\delta$  \textit {vs.} $\ln t$, where  $\delta = \ln
U(t)/\ln  p(t)$,  for  the  $\abmod$  model.  Inset:  blow-up  of  the
right-hand part of the data.}
\label{fig8}
\end{figure}

\noindent Early-time lattice effects, where the different densities do
not overlap, are evident in  Fig.\ \ref{fig7}. It is possible that all
of our data for $p(t)$ and $U(t)$ are within the preasymptotic regime,
since  the  $p(t)$  curve  has  not yet  saturated  to  the  late-time
asymptotic values impied by  Eq.\ (\ref{eq3}).  Note, however, that if
the curves in Fig.\ \ref{fig7} were asympotically linearly increasing,
rather  than eventually  saturating to  a constant,  the corresponding
forms       for      $p(t)$       and       $U(t)$      would       be
$\exp[-C(\rho_0^2Dt)^{1/4}\ln(\rho_0^2Dt)]$,   with  $C$   a  constant
(different  for  $p$  and  $U$).    This  form  differs  only  by  the
logarithmic term from that predicted in \cite{6} on the basis of a toy
model.   It should be  noted that  neither our  present data  nor that
presented  in  \cite{6} can  definitively  rule  out such  logarithmic
corrections,  and the  asympotic form  of $p(t)$,  as well  as $U(t)$,
cannot be  regarded as being definitively  established. Throughout the
preasymptotic  regime,  however,  the relation  $U(t)=K[p(t)]^\delta$,
with $\delta  = 1.39(1)$, holds  rather well.  In Fig.\  \ref{fig8} we
plot the  effective exponent $\delta(t) = \ln  U(t)/\ln p(t)$ directly
as a function of $\ln (\rho_0^2 Dt)$.

\noindent Fig.\  \ref{fig8} is remarkably similar  to Fig.\ \ref{fig4}
for the $\amod$  process.  Indeed, except for the  earliest times, the
corresponding curves differ by less  by $1\%$ everywhere. In a similar
way we can bound the  asymptotic $\delta$ (assuming that the late-time
trend evident in  Fig.\ \ref{fig8} continues) by $1  < \delta < 1.39$.
Since we have  argued elsewhere \cite{6}, using a  toy model, that the
asymptotic  behavior of $P(t)$  is described  by Eq.\  (\ref{eq3}), we
conjecture that the asymptotics of $U(t)$ are given by
\begin{equation}
U(t) \sim \exp[-\delta C(\rho_{0}^{2} D t)^{1/4}]\ ,
\label{eq7}
\end{equation}
where  $C$ is  a constant  and $1  <\delta <  1.39$.  For  the $\amod$
model, we  expressed [Eq.\ (\ref{eq6})]  $U(t)$ in terms of  the ratio
$D'/D$, where $D'$ is the diffusion constant of a tagged particle, $D$
is the  diffusion constant of the  other particles, and  $U(t)$ is the
probability  that the tagged  particle   remains uninfected    at time
$t$. The analogous form for the $\abmod$ process is naturally,
\begin{equation}
U(t) \sim \exp[-(\rho_{0}^{2}Dt)^{1/4}G(D'/D)]
\label{eq8}
\end{equation} 
where  $G(x)$ is  some unknown  function. As  before, the  case $D'=0$
corresponds to  the problem of  unvisited sites, so that  $G(0)=C$ and
$U(t)=P(t)$.   For  $D'=D$, the  case  studied  above, our  conjecture
implies  $G(1) =  \delta C$.  However,  we stress  that a  logarithmic
correction  of the form  discussed above  cannot be  ruled out  by the
data.  The main  motivation  for  Eq.\ (\ref{eq8})  is  the toy  model
presented in \cite{6}, applied to the case $D'=0$.

\section{Discussion and Summary}

In  this   paper  we   have  studied  two   distinct  models:   (i)  a
one-dimensional system of  non-interacting, unbiased diffusing walkers
$\amod$,  with  random  initial   conditions  and  (ii)  the  $\abmod$
diffusion  limited  process  in   one  dimension  with  equal  initial
densities  of  $A$  and  $B$ particles  distributed  homogeneously  at
$t=0$. We have reduced the  familiar study of persistent sites $P(t)$,
to  a limiting  case  in the  study  of uninfected  walkers and  shown
numerically that in both models,
\begin{equation}
U(t) \sim [P(t)]^\gamma\ ,
\label{eq9}
\end{equation}
where $\gamma  \simeq 1.39$ for the  time regimes covered  by our data
and $1  < \gamma  < 1.39$ for  asymptotically large time.   The common
feature of the $\amod$ and  $\abmod$ processes is that the fraction of
unvisited sites decays according  to a stretched exponential (possibly
with  logarithmic corrections  for  $\abmod$).  In  our earlier  paper
\cite{6}, we argued  heuristically, on the basis of  a toy model, that
if the density of walkers decays asymptotically as $t^{-\alpha}$, then
the  fraction  of  unvisited  sites  $P(t)$ decays  with  a  stretched
exponential form if $\alpha < 1/2$. We showed that
\begin{equation}
P(t) \sim \exp-[A(\alpha)\rho_{0}^{2}Dt]^{(1/2)-\alpha}, \qquad \alpha
< 1/2
\label{eq10}
\end{equation}
where $A$  is some  function of $\alpha$.   This result  is consistent
with  the   results  for   models  studied  here.    Remarkably,  Eq.\
(\ref{eq9})  seems to  hold  for  both the  $\amod$  and the  $\abmod$
processes, with approximately the same bounds on $\gamma$. Furthermore,
the  effective time-dependent value  of $\gamma$  is very  similar for
both models  (see Figs.\ \ref{fig4}  and \ref{fig8}). An  obvious goal
for the future is to try to obtain a theoretical understanding of this
simple relationship, and  test it against other models  with $\alpha <
1/2$.

For  both models  studied here  we  expressed the  probability that  a
particle at site $x$ is  uninfected in terms of its diffusion constant
$D'$  and the  diffusion constant  $D$ of  the other  particles [Eqs.\
(\ref{eq6}) and  (\ref{eq8})], noting  that $P(t)$ corresponds  to the
limit  $D'=0$.  Using  Eq.\  (\ref{eq10})  , we  can  generalize  this
description to  any system of  randomly diffusing particles  for which
the  density decays  as $\sim  t^{-\alpha}$  with $\alpha  < 1/2$.  We
conjecture that, in this case,  the fraction of uninfected walkers can
be expressed, asymptotically, as
\begin{equation} 
U(t) \sim \exp[-(\rho_{0}^{2}Dt)^{(1/2)-\alpha}R(D'/D,\alpha)]\ ,
\label{eq12}
\end{equation}
with $U(D'=0)=P(t)$  and $R$ some unknown function.  Clearly, there is
much  scope to  test the  validity  of this  form since  we have  only
considered $\alpha = 0, 1/4$  and $D'=0,D$. Note that for systems with
$\alpha > 1/2$, our toy model predicts \cite{6} that $P(t)$ approaches
a non-zero constant, while  $U(t)$ certainly behaves differently since
it is bounded  above by $\rho(t)$ and must  therefore vanish for large
$t$.   More interesting,  however, is  the borderline  case  where the
particle density  falls off  like $t^{-1/2}$.  In  this case,  our toy
model predicts that $P(t)$ \cite{6}  decays as a power-law, $P(t) \sim
t^{-\theta}$. A  system where  such behavior is  observed is  the well
studied $q$-state Potts model  \cite{8,9,11,12,19}. The results of our
study of uninfected walkers in  this model follows in a separate paper
\cite{32}.

\section{ACKNOWLEDGMENT}
We  thank  Peter Grassberger  for  useful  comments  and pointing  out
references   \cite{20,21,22},   and    Richard   Blythe   for   useful
discussions. This work was supported by EPSRC (UK).

\end{multicols}

\end{document}